\documentclass[9pt, twocolumn]{article}

\usepackage[utf8]{inputenc}
\usepackage{titlesec}
\usepackage{textgreek}
\usepackage[a4paper,top=2cm,bottom=2cm,left=2cm,right=2cm,marginparwidth=1.75cm]{geometry}
\usepackage{amsmath}
\usepackage{float,graphicx}
\usepackage[capitalise]{cleveref} 
\usepackage{xcolor}

\usepackage{widetext}
\usepackage{physics}
\usepackage{siunitx}
\usepackage{authblk}

\newcommand{\qE}{\hat{\mathcal{E}}}

\title{Stationary light in atomic media}
\author[1,2]{Jesse L. Everett}
\author[1,3]{Daniel B. Higginbottom}
\author[1]{Geoff T. Campbell}
\author[1]{Ping~Koy~Lam}
\author[1]{Ben~C.~Buchler}
\affil[1]{Centre for Quantum Computation and Communication Technology, Research School of Physics and Engineering, Australian National University, Canberra, ACT 2601, Australia}
\affil[2]{Light-Matter Interactions for Quantum Technologies Unit, Okinawa Institute of Science and Technology Graduate University, Onna, Okinawa, 904 - 0495, Japan.}
\affil[3]{Department of Physics, Simon Fraser University, Burnaby, British Columbia, Canada V5A 1S6}

\date{}

\begin{document}
\maketitle

\textbf{
 When ensembles of atoms interact with coherent light fields a great many interesting and useful effects can be observed. In particular, the group velocity of the coherent fields can be modified dramatically.  Electromagnetically induced transparency is perhaps the best known example, giving rise to very slow light.  Careful tuning of the optical fields can also produce stored light where a light field is mapped completely into a coherence of the atomic ensemble. In contrast to stored light, in which the optical field is extinguished, stationary light is a bright field of light with a group velocity of zero. Stationary light has applications in situations where it is important to maintain an optical field, such as attempts to engineer large nonlinear interactions.  In this paper we review the stationary light demonstrations published to date and provide a unified theoretical framework that describes the experimental observations. We also discuss possible applications of stationary light with a particular focus on all-optical phase gates for quantum information technology.
}

\section{Introduction}

Several decades of research into coherent atom-light interactions have precipitated a multifarious menagerie of optical phenomena for storing and manipulating light fields inside atomic ensembles \cite{fleischhauer_electromagnetically_2005,Hammerer:2010gsa}. In 2002 Andr\'e and Lukin proposed that dynamically modulating the refractive index along the optical axis of an ensemble could be used, not only to slow or store light, but also to reversibly trap a light field within the atoms \cite{andre_manipulating_2002}. In contrast to prior methods of creating `stored light' \cite{Phillips:2001td}, the optical component of this `stationary light' (SL) field remains considerable even as the light's group velocity vanishes. Rather than mapping the optical field to an entirely atomic state, the non-zero light field is prevented from propagating by a dynamic optical bandgap, analogous to the static bandgap caused by the structure of photonic crystals or ordered atoms \cite{schilke_photonic_2011}. This bandgap is controlled by counterpropagating optical fields and can be tuned to manipulate the localization of light fields and atomic excitations inside the ensemble.

The original SL proposal \cite{andre_manipulating_2002} was followed in quick succession by a first experimental demonstration in hot atomic vapor \cite{bajcsy_stationary_2003} and alternative SL schemes, some with an alternative physical picture of the underlying mechanism \cite{moiseev_quantum_2006}.  Although early demonstrations were described in terms of standing wave modulated gratings, this picture does not apply to hot atoms due to thermal atomic motion. A subsequent multi-wave mixing formulation \cite{moiseev_quantum_2006,zimmer_coherent_2006} more comprehensively explains the complex behaviours resulting from combinations of counterpropagating optical fields.

The same all-optical tunable bandgap that lies behind SL effects has been considered as a flexible alternative to fixed photonic crystals \cite{artoni_optically_2006} with  applications in quantum light storage and fast optical switching. Furthermore, because of the nonlinear behaviour of atomic ensembles, reversibly trapping a SL field in such a dynamic bandgap holds promise for enhancing nonlinear photon-photon interactions.  The use of SL for this purpose is strongly motivated by the development of photonic quantum information processing \cite{chuang_simple_1995}. The size of a nonlinear phase shift scales with the product of the interaction strength and time. Consequently, nonlinear interactions usually involve high intensity fields. Photonic qubit gates, however, require nonlinear interactions for fields down to the single-photon level. SL therefore provides a path to this end by localizing optical fields in the atomic medium and providing a longer time for a nonlinear phase shift to be accumulated.

Although slow-light schemes have been proposed for such quantum information applications, they feature an inherent trade-off between interaction time and interaction strength, because the photonic component of the polariton is inversely proportional to the time the probe spends inside the ensemble \cite{harris_nonlinear_1999}. SL schemes allow greater flexibility in the configuration of the optical field potentially enabling larger conditional phase shifts at the single photon level \cite{andre_nonlinear_2005} and photon-photon entanglement \cite{friedler_deterministic_2005}.

These SL phase-gate schemes are in some ways analogous to the well known cavity QED techniques for atom-mediated photon-photon gates \cite{obrien_2009_nat_phot}, with the photonic band gap trapping the light field in place of an optical resonator. What distinguishes SL photon-photon gate proposals is the wide degree of tunability: the spatial distribution of stationary atomic coherences and optical fields can be separately configured to implement a wide range of potential interactions.

The purpose of this review is to consolidate the new body of work on SL and to provide a unified model of SL phenomena under various conditions given the experimental evidence now available. We will also consider the prospects and limitations of SL as a tool for quantum information applications in light of these results.

\subsection{Structure}

We have divided the following review into three sections. We begin in \Cref{sec:gen} with the literature concerning schemes for generating SL fields. Our intention in reviewing this work is to convey a sense of the history of the field, with an emphasis on experimental results and how they shaped our evolving picture of SL. We divide these results by generation scheme. The bulk of results to date concern EIT-based SL, which we cover in \Cref{sec:EITSL}. The more recent Raman-based schemes are covered in \Cref{sec:ORSL}. In the interests of brevity, we will at first introduce only the bare minimum theory required to provide perspective for these results.

In \Cref{sec:theo} we provide a mathematical basis for the physics in this review and introduce a comprehensive theoretical framework for SL in atomic ensembles. Our goal is to give the reader sufficient tools to explain and model the effects discussed in \Cref{sec:gen} as well as proposals which have yet to be implemented. We derive equations of motion for the optical fields and atomic coherences in a secular level scheme, i.e. one in which only the interactions of copropagating fields are considered. We use these to describe EIT and Raman SL. We return to a non-secular scheme to discuss how to calculate the effect of higher-order coherences on the SL. Finally, we discuss phase-matching requirements and transverse propagation.

\Cref{sec:apps} concerns the proposed uses of SL, in particular for mediating gates in photonic quantum information systems. We review early proposals for enhancing nonlinear interactions, discuss the no-go theorems these proposals inspired, and finally review more recent proposals that should overcome the obstacles raised by the no-go theorems.

\par
\medskip

\section{Generation of stationary light}\label{sec:gen} 
The SL schemes proposed and implemented to date can be divided into two broad categories delineated by the configuration of the optical control generating the stationary field. The earliest SL schemes were based on electromagnetically induced transparency (EIT) with near-resonant control fields and subsequent research has largely focused on this approach. More recently, Raman SL schemes with far-detuned control fields have been introduced.
What follows in this section is a short summary of SL work to date in each of these approaches. 

Although SL can, in principle, be generated in any sufficiently large ensemble of emitters, current demonstrations have been restricted to atomic systems by the difficulty of constructing optically deep and sufficiently coherent ensembles of `artificial atoms' such as quantum dots and diamond colour centres. In particular, most demonstrations have been done in vapours of rubidium or cesium atoms, which are dense ensembles with hydrogen-like spectra. SL has yet to be demonstrated in other optically deep atomic systems, such as rare-Earth ion crystals. Throughout this review we'll refer to the ensemble mediating SL as `atoms', but it's worth bearing in mind that SL could be generated in any optically dense ensemble of coherent emitters.

We'll see in the sections below that SL generation depends sensitively on the mobility of atoms within the ensemble, with qualitatively different behaviour in hot ($\approx 400$~K), cold ($\approx$~mK), and ultracold ($\approx$~$\mu$K) or stationary atoms. Hot, mobile atoms are typically the more complicated platform for quantum optics. Their velocity with respect to optical beams Doppler broadens transitions and local coherences and excitations are carried with the atoms as they move ballistically or diffusively through the ensemble. In this case, however, motion has the effect of simplifying SL generation. Ultracold and stationary atoms can maintain a richer variety of coherences (higher-order coherences, see \Cref{sec:hocs}) and have the more complicated dynamics.

% EIT dispersion figure
\begin{figure*}[th]
\centerline{\includegraphics[width=175mm]{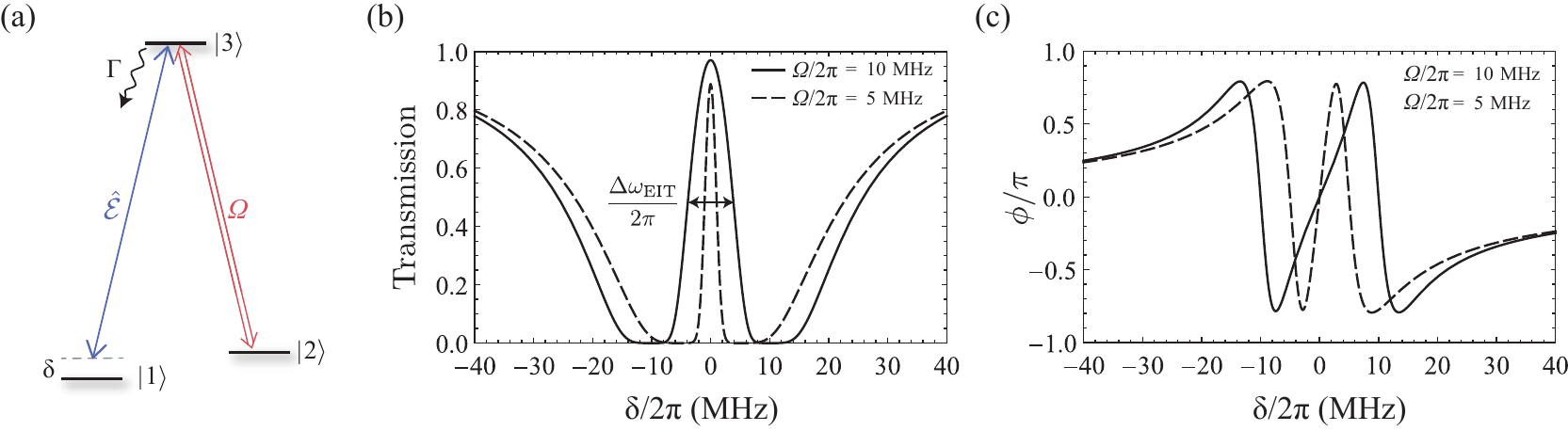}} 
\caption{  \small $\Lambda$ atomic level configuration for electromagnetically induced transparency. A weak probe, $\hat{\mathcal{E}}$, copropagates with a bright control field $\Omega$ which opens a narrow transparency window for the probe about the two-photon resonance $\delta = 0 $. The excited state decays at a rate $\Gamma$. (b) Probe transmission and (c) phase shift, $\phi$ through an ensemble of $\Lambda$-type atoms as a function of two-photon detuning $\delta$. The width of the EIT window, $\Delta \omega_{\mathrm{EIT}}$, increases with control field power (dashed: lower power, solid: higher power).}
\label{fig:eit}
\end{figure*}

\subsection{EIT-based stationary light}\label{sec:EITSL}

Electromagnetically induced transparency is a property of three-level atoms interacting with two (usually copropagating) optical fields. In the most common configuration, known as the $\Lambda$ configuration and shown in \Cref{fig:eit}(a), a weak probe field couples the atomic ground state $\ket{1}$ to an excited state $\ket{3}$ and a bright control beam couples the excited state to a meta-stable state $\ket{2}$. The bright control field, with Rabi frequency $\Omega$, opens a narrow transparency window for the probe, $\qE$, at a resonant frequency that would otherwise be absorbed. When the two fields are equally detuned from the excited state, they are said to be in two-photon resonance and drive the ground-metastable state coherence. When the two-photon resonance condition is exactly met, the probe and control field excitations interfere such that the excited state is not driven at all, and the atomic medium is rendered transparent for the probe field. This is known as electromagnetically induced transparency (EIT) and can be used to open a narrow window of almost perfect transparency in an otherwise opaque atomic ensemble. The bandwidth of the transparency window, shown in \Cref{fig:eit}(b), is \cite{fleischhauer_quantum_2002}
\begin{align}
\Delta \omega_{\mathrm{EIT}} = \frac{\Omega^2}{\Gamma \sqrt{\alpha}} \,,
\end{align}
where $\Gamma$ is the total decay rate of the excited state $\ket{3}$ and $\alpha$ is opacity in the absence of EIT.

In an ensemble of emitters, EIT gives rise to a controllable slow-light effect for the probe field. The group velocity of the probe is proportional to the phase/frequency gradient $\partial \phi / \partial \delta$, shown in \Cref{fig:eit}(c), which is inversely proportional to the linewidth of the transparency window. The width of the transparency is, in turn, proportional to the power of the control field, providing an adjustable group velocity. EIT has been used to slow classical light pulses down to $17$~m/s \cite{hau_light_1999} and single photons to $10^3$~km/s \cite{eisaman_electromagnetically_2005}.   

The slow light in EIT exists as a polariton superposition of an optical field and atomic coherence. The coherence is frequently generated between hyperfine split ground states in which case we may call the coherence envelope a `spinwave'. The greater the atomic proportion of the polariton, the slower it propagates through the ensemble. By adiabatically reducing the power in the control field, a resonant probe field can be decelerated. As the light slows, the optical component of the polarition is reduced while the spinwave component grows. When the control field power reaches zero, the probe field  becomes a state of  `stored light' that has no optical component. The stored light spinwave is motionless, which is why it is also sometimes referred to as `stopped light'.

EIT has been proposed as a means to create a memory device for quantum light \cite{fleischhauer_quantum_2002,fleischhauer_electromagnetically_2005} with applications in quantum communication \cite{lvovsky2009optical}. Experimental demonstrations have shown recall of non-classical states of light \cite{akamatsu_squeezed_2004, appel2008quantum, chaneliere2005storage} and efficiencies of up to 92\% \cite{PhysRevLett.120.183602}. EIT has also been proposed to enhance non-linearities for all-optical quantum gates \cite{schmidt1996giant, fleischhauer_electromagnetically_2005, chang2014quantum}. Further uses for EIT include slow-light enhanced sensing \cite{purves2006sagnac} and laser cooling below the Doppler limit \cite{roos2000experimental}.

\begin{figure*}[t]
\centerline{\includegraphics[width=150mm]{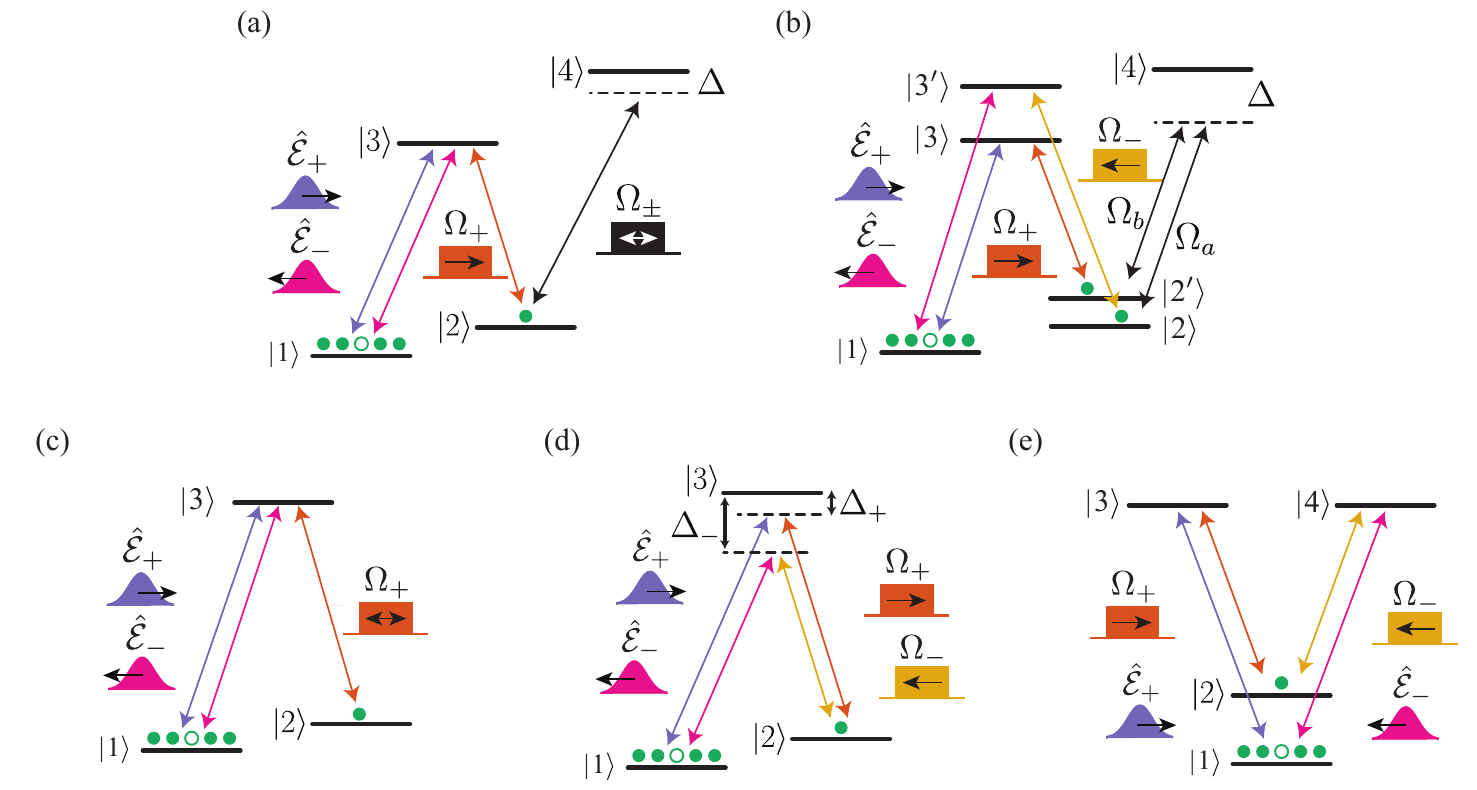}} 
\caption{  \small EIT SL level schemes. (a) Original proposal by Andr\'e and Lukin for EIT SL in stationary atoms \cite{andre_manipulating_2002}. (b) Multi-colour scheme proposed by Andr\'e and Lukin as a Doppler-free alternative to (a) for use in hot-atom ensembles \cite{andre_manipulating_2002}. (c) Scheme used by Bajcsy et al. in the first demonstration of SL \cite{bajcsy_stationary_2003}. When performed in a medium of stationary atoms, this scheme drives the creation of higher-order coherences (HOCs). (d) Two-colour SL scheme with a single excited state. Higher order coherences may arise depending on the temperature and choice of detunings $\Delta_\pm$. (e) Secular SL scheme with separate excited states addressed by the counterpropagating fields. This prevents the creation of HOCs, and is functionally equivalent to (c) in a hot-atom where atomic motion washes out HOCs. This is the secular approximation for hot atoms.}
\label{fig:schemes}
\end{figure*}

\subsubsection{Proposal and early demonstrations}

\begin{figure}[b!]
\centerline{\includegraphics[width=85mm]{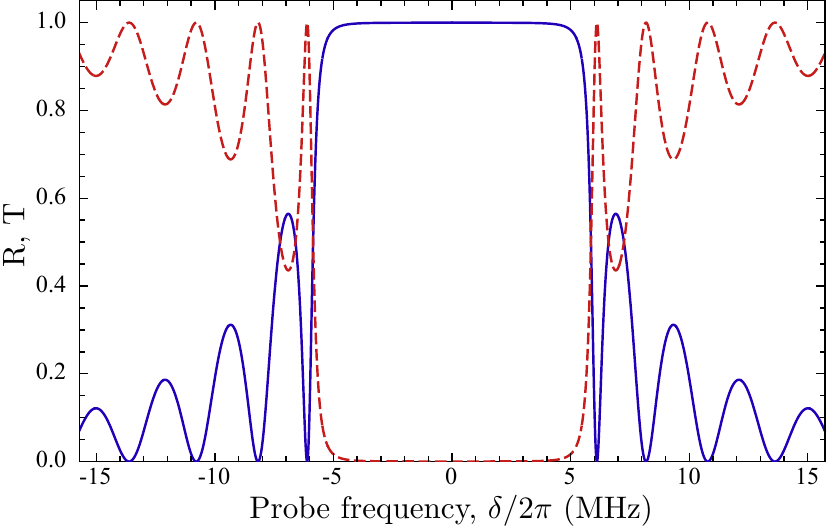}} 
\caption{\small The transmission (dashed red line) and reflection (solid blue line) of a probe field incident on an atomic ensemble with periodically modulated dispersion  (using the model in \cite{lahad_induced_2017}). A bandgap emerges on resonance, preventing the propagation of light within the ensemble while light incident on the ensemble is reflected. Off resonance, transmission peaks appear as interference between light reflected throughout the ensemble becomes constructive at the far end.}
\label{fig:bandgap}
\end{figure}

The first SL proposal by Andr\'e and Lukin \cite{andre_manipulating_2002} involved applying a spatially modulated light-shift to an EIT window via an additional optical standing wave addressing the metastable state and a fourth level as illustrated in \Cref{fig:schemes}(a). This was identified as creating a dispersive Bragg grating by periodically modulating the EIT transparency frequency across the ensemble, trapping the light. An example of a SL-induced bandgap is shown in \Cref{fig:bandgap}.  

It was already recognized in this initial proposal that atomic motion would effect the SL dynamics via Doppler shifting. In the conclusion to Ref.~\cite{andre_manipulating_2002}, Andr\'e and Lukin proposed a Doppler-free alternative for generating SL in hot atoms. This scheme, shown in \Cref{fig:schemes}(b), couples the forward and backward propagating probes to separate coherences between distinct metastable states. In contrast to the scheme of \Cref{fig:schemes}(a), this alternative proposal is `multi-colour' in that the counterpropagating probe and control fields are at different frequencies.      

The first demonstration of SL by Bajcsy et al. followed soon after this proposal, and took a simpler but related approach \cite{bajcsy_stationary_2003}. Rather than modulate the EIT frequency with an additional off-resonant standing wave, as in \Cref{fig:schemes}(a), a standing wave was generated in the control field intensity itself, see \Cref{fig:schemes}(c). This was thought to produce a standing wave in the probe absorption due to the spatial modulation of the EIT effect (an electromagnetically induced grating, or EIG \cite{brown_all-optical_2005}) and result in a stationary probe field with intensity fringes at the control field wavelength. SL fields  generated by this approach in a hot Rubidium vapour cell were witnessed by controllably releasing the field from the ensemble.

The SL bandgap was further investigated by Brown et al. by measuring the reflection of a probe field from a hot $^{87}$Rb vapour cell illuminated by counterpropagating control fields \cite{brown_all-optical_2005} using the scheme shown in \Cref{fig:schemes}(c). In this experiment the reflected power was limited to less than $7$\% owing to the small fraction of atoms slow enough to contribute to a grating. Performance may also have been limited by phase mismatch between the forward and backward control fields.

A similar experiment \cite{wang_optical_2013} was carried out more recently in room temperature Cs and showed that the SL bandgap can be made direction sensitive by detuning the counterpropagating control fields as in \Cref{fig:schemes}(d). In this experiment the detuning  ($\Delta_{+}-\Delta_{-}$) was 20~MHz. The result was described as an ``all optical diode'' and was explained in terms of a  ``travelling photonic crystal''. The standing wave between the two counterpropagating control fields $\Omega_\pm$ travels with velocity proportional to their frequency difference $\Delta_+ - \Delta_-$. In the frame of the traveling grating, forward and backward propagating probe fields are blue and red Doppler shifted respectively. By choosing detunings such that only backward propagating fields fall into the bandgap, the medium becomes direction sensitive. In this experiment reflectance approached unity, considerably higher than the earlier work of Ref.~\cite{brown_all-optical_2005} even in a hot-atom medium. Ref.~\cite{ullah_observation_2014} added a fourth field in order to observe the interplay of EIT SL bandgap with non-linear parametric field generation by four-wave mixing.

Although the first two demonstrations, Refs.~\cite{bajcsy_stationary_2003, brown_all-optical_2005}, would be the only two experimental investigations of SL for four more years, a great deal of theoretical work was carried out in this time.  In particular, it was realized that the intuitive model of SL arising from a standing wave of the control field does not work when one considers the motion of atomic vapours. The motion of the atoms across a period of the standing wave modulating the EIT window is much faster than the speed of the EIT interaction itself, as given by the EIT inverse bandwidth $\tau_\mathrm{EIT} = 1/ \Delta \omega_\mathrm{EIT} $. In this case probe light travelling through the medium does not actually experience a spatially modulated absorption or dispersion profile. 

Hot atoms travel through alternate regions of high and low control field intensity during the EIT process, and so the standing wave grating is insufficient to explain how a stationary probe field is generated in the above experiments. Alternative schemes were also proposed for generating SL (in hot and ultracold atomic media) in which the standing wave grating picture would prove inadequate. A complete treatment of these many SL schemes requires considering multi-wave mixing, the coupling of coherences to counterpropagating fields, which is outlined in the following section.

\subsubsection{Multi-wave mixing theory}
Multi-wave mixing (MWM) is a process where the ground state atomic coherence is coupled to both forward and backward travelling probe fields by their respective control fields. Under control field parameters that give SL conditions, the resulting polariton has zero group velocity. The interpretation for the SL in this case is that the polariton is prevented from spreading by the interference between light travelling along multiple different paths. That is, the light is reflected at a continuum of different points in space, generating interference. This is equivalent to the mechanism by which a Bragg (absorption) grating produces a bandgap. Although they have similar consequences in the simplest configuration, the multi-wave mixing description is necessary to account for the interactions that generate higher-order coherences (HOCs). 

The multi-wave mixing treatment of SL began with Moiseev and Ham, who considered several situations in which SL is generated without a standing wave control field. They showed that a SL field could be generated from an EIT slow light pulse by adiabatically switching on the counterpropagating control field \cite{moiseev_generation_2005}, and that forward and backward control fields of two \cite{moiseev_quantum_2006} or more \cite{moiseev_quantum_2007} different frequencies could generate multi-colour SL fields. The bichromatic control field scheme shown in \Cref{fig:schemes}(d) generates SL fields even when the mutual frequency difference is so large that no control field standing wave exists (in contrast to the small frequency difference used for the optical diode \cite{wang_optical_2013}). Moiseev and Ham explored the use of this scheme for wavelength conversion by adiabatically switching off the forward control field \cite{moiseev_quantum_2006}.

Zimmer et al. gave a detailed explanation of SL in a hot-atom medium based on multi-wave mixing \cite{zimmer_coherent_2006}. The authors drew upon pulse matching as described for EIT \cite{Harris1993} to derive the characteristic spreading time of the quasi-stationary probe pulse expanding to match the stationary control profile in a medium with finite optical depth.

\subsubsection{High-order coherences}\label{sec:hocs}

Coupling between forward and backward propagating fields in the multi-wave mixing theory generates coherences with higher momentum than the probe and control photons, as it involves the absorption and re-emission of photons traveling in opposite directions. These high momentum spinwaves are known as higher-order coherences (HOCs) and have a wavelength of half the optical wavelength or smaller. We use the term `coherence’ as HOCs include both spinwaves (ground state coherences) and excited state coherences. The coherence order refers to the additional momentum, for example a $+2$ order coherence is generated by absorption of a forward probe photon with re-emission into the backward control field. Higher orders are generated by the subsequent absorption and re-emission of additional counterpropagating fields.

The movement of hot atoms (under the experimental conditions of Ref.~\cite{bajcsy_stationary_2003} and other demonstrations in hot atoms) washes out the sub-wavelength spinwave more quickly than the light can couple to it. Under such conditions HOCs are not important, and simply decohere before they can be coupled. In contrast, this washing out occurs much more slowly in cold or stationary atoms and these HOCs can indeed effect the propagation of light. 

The case of EIT with frequency degenerate counterpropagating control fields (\Cref{fig:schemes}(c)) in stationary atoms was considered by Hansen et al.\cite{hansen_trapping_2007}, but only in the `secular' approximation in which coherences couple exclusively to copropagating fields and cross-coupling between counterpropagating fields is forbidden. Because such cross-coupling is the origin of HOCs, HOC formation is prevented in this system. This secular approximation is made exact in the dual-V level scheme shown in \Cref{fig:schemes}(e). The authors conclude that SL is possible in stationary atoms with such a scheme. In a hot-atom medium that cannot sustain HOCs, the classic single-colour EIT SL configuration of Ref.~\cite{bajcsy_stationary_2003} and \Cref{fig:schemes}(c) is equivalent to the Doppler-free, secular scheme of \Cref{fig:schemes}(e).

The standing wave grating description of the SL in Ref.~\cite{bajcsy_stationary_2003} is mathematically equivalent to a description of multi-wave mixing in the secular regime. This isn't surprising, as either mechanism creates a coherent interchange of forward travelling light with backward travelling light, resulting in the same bandgap behaviour. Multi-wave mixing can give rise to a bandgap with the same profile as the dispersive grating bandgap in  \Cref{fig:bandgap}. 

The generation of HOCs is first mentioned by Moiseev and Ham \cite{moiseev_generation_2005}, and thoroughly examined by Hansen and Mølmer \cite{hansen_trapping_2007,hansen_stationary_2007}. In particular, the later authors examine how HOCs can negatively affect the generation of SL. The additional interference between light generated from these various HOCs can disturb the multi-wave mixing SL effect \cite{hansen_stationary_2007, wu_stationary_2010}. Under these conditions the polariton can split and travel through the ensemble, and no SL field exists.

\subsubsection{Demonstrations with cold atoms}

\begin{figure}[t!]
\centerline{\includegraphics[width=80mm]{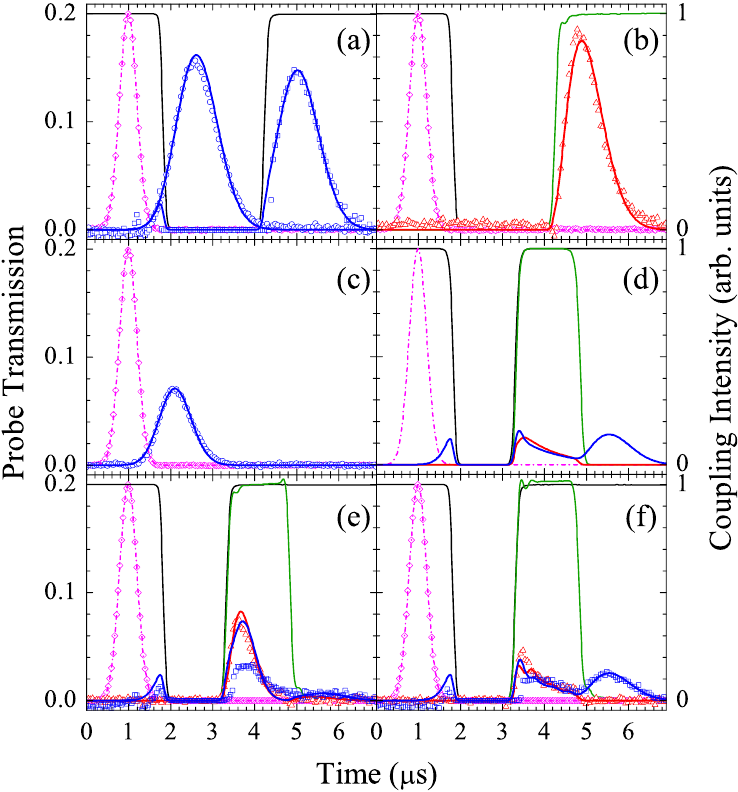}} 
\caption{\small  Propagation of a pulse (pink diamonds, scaled by 0.2) through an ultracold atomic medium with combinations of slow, stored and SL. Experiments (points) are compared to theoretical predictions (lines). Control field intensity is shown in black (forward) and green (backwards). Recalled probe intensity is shown in blue (forwards) and red (backwards).
(a) Forward probe delayed by slow light (circles, continuous control not shown) and retrieved after having been converted to stored light (squares, switched control shown).
(b) Backward probe retrieved with backward control.
(c) Forward probe delayed by continuous backward control (not shown).
(d-f) show a sequence of stored light (2-3$\mu s$) followed by SL (3-5$\mu s$) and forward recall (5$\mu s$).
(d) A model for a hot atomic medium with single-colour control fields.
(e) Model and experiment for ultracold atomic vapour with single-colour control fields. Substantial probe light leaks both forward and backward during the SL stage. The absence of retrieved probe compared to the hot-atom case indicates that SL was unsuccessful.
(f) Mutually-detuned (two-colour) control fields and ultracold atoms. The smaller leakage of the probe and presence of recalled forward probe indicates the successful creation of SL. (Figure modified from \cite{lin_stationary_2009}).}
\label{fig:lin2009}
\end{figure}

A convincing demonstration of the role of HOCs in the multi-wave mixing process was presented by Lin et al.~\cite{lin_stationary_2009}. This work compared the schemes of \Cref{fig:schemes} (c) and (e) and  the key results are shown in \Cref{fig:lin2009}. This work demonstrated that no SL field is formed in an ultracold-atom medium when the counterpropagating EIT control fields are frequency degenerate, \Cref{fig:lin2009}(e). This is consistent with the destructive role of HOCs introduced in the multi-wave mixing theory~\cite{hansen_stationary_2007}. 
When using a scheme where the forward and backward control fields had very different frequencies, Ref.~\cite{lin_stationary_2009} successfully demonstrated the formation of SL  \Cref{fig:lin2009}(f). This was the first experiment to generate SL fields in an ensemble of laser-cooled atoms. The forward propagating probe and control pair had a wavelength of $780$~nm and the backward pair $795$~nm. Due to the wavelength difference, the subsequent two-photon travelling wave modulates the control field intensity grating several orders of magnitude more quickly than the EIT bandwidth. The energy difference also prevented direct coupling between the forward probe and backward control and vice versa. The SL field is therefore generated without any standing wave in the control field. The large difference in wavelengths ensured that the HOCs are suppressed, but also prevented effective phase-matching of the counterpropagating fields, causing rapid decay of the spinwave during the SL period.

In their analysis, Lin et al. matched the shape of of leaked and released SL fields to a multi-wave mixing model that included excited state coherences of order $\pm1$ driving spin wave coherences of order $\pm 2$ with qualitative agreement. Subsequent work by Wu et al. \cite{wu_stationary_2010} expanded on this multi-wave mixing model by including both higher order coherences and residual Doppler broadening, which remains non-negligible at temperatures of several hundred $\mu$K and was included in an earlier theoretical treatment by the same authors \cite{wu_decay_2010}. The higher order, Doppler broadened MWM model matches additional features of the data taken by Lin et al. in Ref.~\cite{lin_stationary_2009}.  

In a pair of follow-up experiments \cite{peters_observation_2010, peters_formation_2012} Peters et al. further explored SL formation in ultracold ensembles with the same apparatus as Ref.~\cite{lin_stationary_2009}. First they prepared SL using a much smaller forward and backwards detuning difference ($\Delta_+ - \Delta_-$) on the order of $10$~MHz \cite{peters_observation_2010}. In this configuration the off-resonant transitions are weakly driven, but may nevertheless introduce a non-uniform, time varying phase variation to the SL field (a prediction made by Moiseev and Ham in their MWM theory \cite{moiseev_quantum_2006}). Peters et al. inferred the SL phase shift by inference from a numerical MWM model matched to the phase of released SL fields as predicted by a numerical MWM model. This phase variation poses a problem for quantum information schemes that propose to use cross-phase modulation from the SL field to phase shift a target pulse (which we discuss in \Cref{sec:apps}). This phase distortion is, however, not fatal for SL cross-phase modulation with cold atoms because the distortion vanishes for large ODs.

In Ref.~\cite{peters_formation_2012} Peters et al. considered the role of EIT window bandwidth in the formation of SL pulses in ultracold media, showing that HOCs detrimental to SL are formed when the EIT bandwidth is larger than the ensemble Doppler broadening. This condition expresses the ability for mobile atoms to sustain wavelength-scale coherences. 

Although HOCs can prove destructive for SL fields, Park et al. showed that the frustrated SL effect could be used to construct a coherent and dynamic beam splitter with a cold vapour of magneto-optically trapped rubidium atoms \cite{park_coherent_2016}. MWM by counterpropagating control fields coherently couples a stored spinwave into two optical modes with power splitting ratio, phase and frequency determined dynamically by the controls. Such an effect would not be possible in a hot medium that cannot sustain HOCs.

\subsubsection{Stationary light in hollow core fibre}

The optically deep cold-atom ensembles discussed so far were realized by using a magneto-optical trap (MOT) to cool and confine a cloud of atomic vapour within a vacuum chamber. In Ref.~\cite{blatt_stationary_2016} Blatt et al. demonstrated SL formation with EIT in an atomic ensemble inside a hollow-core using the scheme of \Cref{fig:schemes}(d) with detunings large enough to suppress HOCs. The fiber was loaded with cold rubidium atoms from a MOT, at a temperature of $1$~mK. SL was witnessed by the suppression of an EIT pulse escaping from the ensemble. Once again, SL could be maintained only when the relative detuning of the two counterpropagating control fields was larger than the EIT transmission bandwidth.   

In such hollow-core fibre experiments, the fields propagate along a single optical axis. This limits the capacity to phase-match the counterpropagating fields, which can be done in free-space systems by introducing small angles between the fields (see \Cref{sec:phasematch}). In Ref.~\cite{blatt_stationary_2016} this was compensated with a combination of two-photon detuning $\delta$ and an imbalance between the two control field Rabi frequencies $\Omega_{c\pm}$. 

Loading atoms into hollow core fibers has the consequence of increasing the atom-photon interaction cross section. The optical field is confined largely within the fibre core so that the field per photon remains large as the guided mode propagates across the ensemble. This is particularly advantageous for SL cross-phase modulation schemes which improve not only with high optical depths, but also with the single photon-single atom interaction cross section \cite{hafezi_quantum_2012}.

Fibers also bring additional challenges due to the confined geometry. Firstly there is the issue of atomic collisions with the walls of the hollow fiber. One approach to mitigate this issue is to coat the inside of the fiber with an anti-relaxation coating that reduces collisional dephasing. Using room temperature atoms this approach has yielded a collisional dephasing rate of $\sim$1~MHz \cite{light_OL_2007}. A dipole trap can also be used to trap the atoms in the center of the fiber core, away from the walls. The SL demonstration in Ref.~\cite{blatt_stationary_2016} used a red-detuned dipole trap with a depth of $5$~mK resulting a collisional dephasing rate of only 50kHz. A second issue is control field  inhomogeneity.  In free space experiments, the control fields can be expanded to improve the uniformity of the intensity over the interaction volume. In a fiber this is not possible and radial control field intensity variations impose additional inhomogeneous broadening on the atomic ensemble \cite{blatt_stationary_2016}.

\subsubsection{Imaging the atomic coherence}

\begin{figure*}[th]
\centerline{\includegraphics[width=175mm]{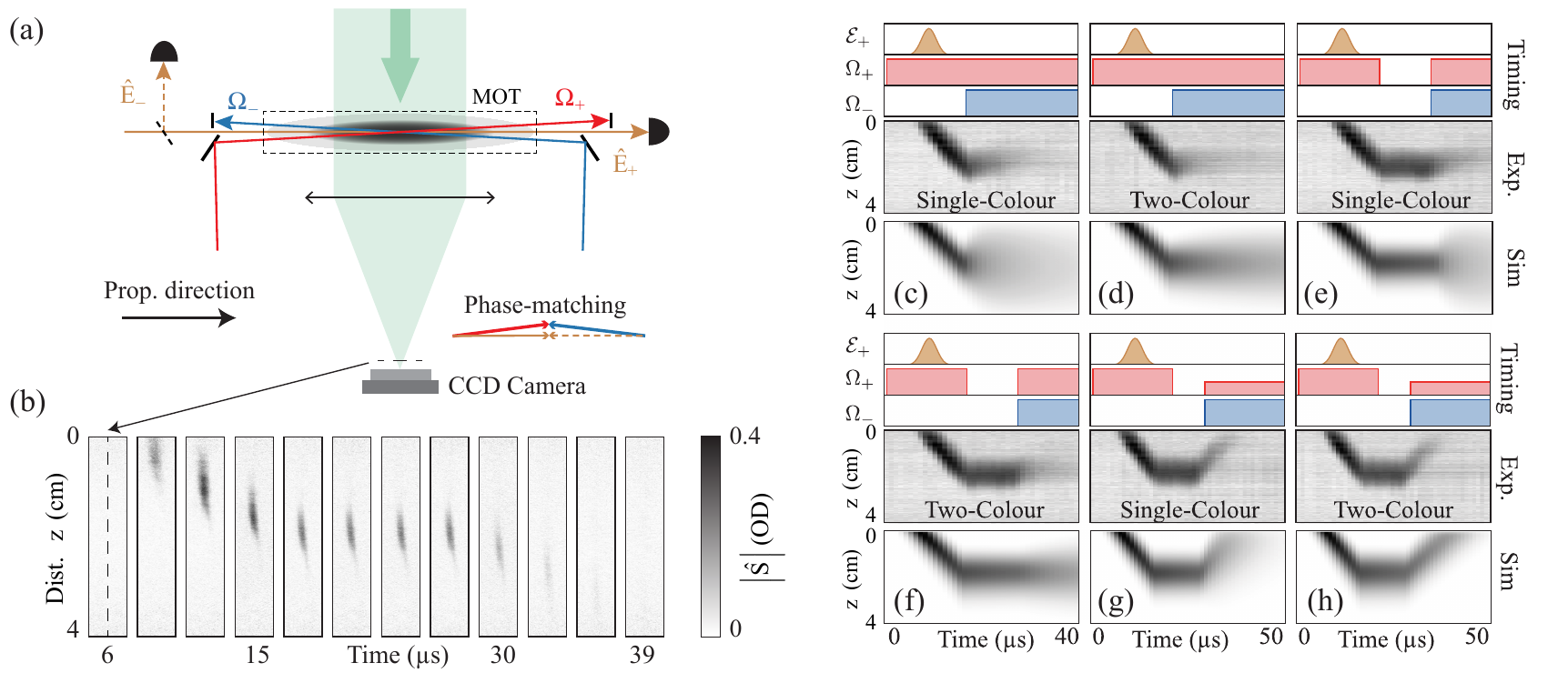}} 
\caption{\small (Adapted from \cite{campbell_direct_2017}). Side-imaging the atomic coherence during EIT SL. (a) Schematic of the apparatus used in Ref.~\cite{campbell_direct_2017} to side-image the atomic coherence of a propagating polariton. An additional imaging beam transversely illuminates the ensemble. (b) Stroboscopic images of shadows cast by an EIT polariton spinwave $\hat{\mathcal{S}}$ propagating through the ensemble. (c--h) Observed evolution of the coherence along the propagation direction compared to MWM simulations. The corresponding pulse scheme is shown at top. The EIT polariton first propagates into the ensemble and is then either stored (`stored light', no control fields) or frozen by EIT SL with either single- or two-colour control fields. (c,d) Balanced EIT SL. (e,f) Stored light followed by EIT SL. (g,h) Unbalanced EIT SL.}   
\label{fig:EITSL_imaging}
\end{figure*}

All the experiments discussed so far inferred the existence of EIT SL by observing the temporal shape of the output optical fields, such as the results in \Cref{fig:lin2009} from Ref.~\cite{lin_stationary_2009}. Optical probe pulses were observed to be reflected by the SL bandgap, recalled from the ensemble after being trapped by the SL bandgap, or leaked forwards and backwards from the ensemble during imperfect SL operation. Significant conclusions about the SL process have been reached by comparing the behaviour of these output fields with models. Throughout these experiments, however, the internal dynamics of the atomic ensemble and its associated light fields remained unobserved.

To expose the internal dynamics of an atomic ensemble, side imaging can be employed. In this measurement a broad imaging beam illuminates the entire ensemble from a direction perpendicular to the optical propagation axis and is absorbed selectively by one of the ground states in the atomic coherence, $\hat{\mathcal{S}}$. An example of this configuration is shown in \Cref{fig:EITSL_imaging}(a). The ensemble's shadow  is imaged onto a camera to reveal the shape of the spinwave. Side imaging the spinwave after various delays reveals the propagation of a polariton through the ensemble stroboscopically by imaging its atomic component at different times. The stroboscopic evolution of an EIT polariton recorded in Ref.~\cite{campbell_direct_2017} is shown in \Cref{fig:EITSL_imaging}(b). The measurements are destructive, so each image represents a new run of the experiment with a fresh ensemble of atoms.

Side imaging had previously been used to observe the propagation of atomic coherences in Bose-Einstein condensates \cite{ginsberg_coherent_2007,zhang_creation_2009} and hot atomic vapours \cite{wilson_slow_2017}. The use of side imaging to expose the dynamics of SL was first demonstrated by Everett et al. \cite{everett_dynamical_2016}.  This experiment used Raman SL and showed very different behaviour to the EIT SL considered here, as will be discussed in detail in \Cref{sec:ORSL}.

Side imaging was first applied to EIT SL in Ref. \cite{campbell_direct_2017}. In this experiment the atomic coherence in an ensemble of cold, magneto-optically trapped atoms was imaged during both slow and SL scenarios. This was the first time that a direct comparison could be made between the actual and simulated evolution of an EIT SL polariton within the ensemble. Since there is little prospect of directly measuring the SL field (if it were observed then it can not also be stationary!) the ability to image the spinwave $\hat{\mathcal{S}}$ and compare this with models is the next best thing.

The experiments in Ref.~\cite{campbell_direct_2017} showed the EIT SL polariton diffusing due to the limited optical depth of the ensemble (\Cref{fig:EITSL_imaging}(c,d)); the motion of the polariton with unbalanced control fields  (\Cref{fig:EITSL_imaging}(g,h)) and compared single-colour (\Cref{fig:schemes}(c)) two-colour SL (\Cref{fig:schemes}(d)) with a relative detuning of $4$~MHz between the forward and backward propagating components (\Cref{fig:EITSL_imaging}(c--f)). In this experiment there was no measurable difference between the single- and two-colour EIT SL. It was supposed that the pump process that prepared the ensemble also induced sufficient longitudinal atomic motion to wash out the control field standing wave.  

\subsubsection{Stationary light with quantum fields}

SL-based gates for photonic quantum information require the trapping of quantum light fields, in many proposals this means generating a SL field with a single photon. To date, however, almost all SL experiments have been done with classical fields.  The first, and so far only, demonstration of a quantum SL field was performed by Park et al. \cite{park_experimental_2018} using single photon states. They generated a single distributed atomic excitation in a cold $^{87}$Rb MOT by detecting a Stokes photon Raman scattered spontaneously from a detuned `write' pulse. Measuring the Stokes photon heralds the existence of a single-excitation spinwave throughout the ensemble.

Counterpropagating control fields, mutually detuned to prevent the creation of HOCs (as in \Cref{fig:schemes}(d)), transfer the single-excitation spinwave to an anti-Stokes single-photon SL field trapped in the ensemble. In contrast to the coherent SL fields in the previous experiments, which were generated from polariton pulses injected into the ensemble, the single-excitation spinwave envelope is uniform because every atom in the ensemble is equally likely to have scattered the Stokes photon. SL can't be maintained at the edges of the ensemble, so a portion of the single-photon field escapes during the SL stage.

The anti-Stokes photon is eventually released by a single-directional control field after being held briefly as a SL pulse. The post-SL anti-Stokes photon is non-classically correlated with the herald Stokes photon, as well as being anti-bunched (under some assumptions) \cite{park_experimental_2018}. Such a demonstration of quantum SL fields is a necessary precursor to photonic computation with SL-mediated interactions.

\subsection{Raman stationary light}\label{sec:ORSL}

The EIT-based form of SL from Andr\'e and Lukin's first proposal relies on a local multi-wave interference mechanism between near-resonant fields. The propagation of light is prevented due to  interference between counterpropagating light fields generated by multi-wave mixing. In contrast, Raman SL uses an interaction in which the driving fields are far detuned from atomic resonance. This form of SL was proposed and demonstrated by Everett et al. in Ref.~\cite{everett_dynamical_2016}. The absorption length of such far-detuned fields is much larger and the interference between light reflected after travelling short distances can therefore not be relied on to trap the light. Instead, the effect that traps Raman SL is interference of light that is generated at different positions in the memory. For this reason, Raman SL can potentially be sustained over larger distances.

\begin{figure*}[thp]
\centerline{\includegraphics[width=160mm]{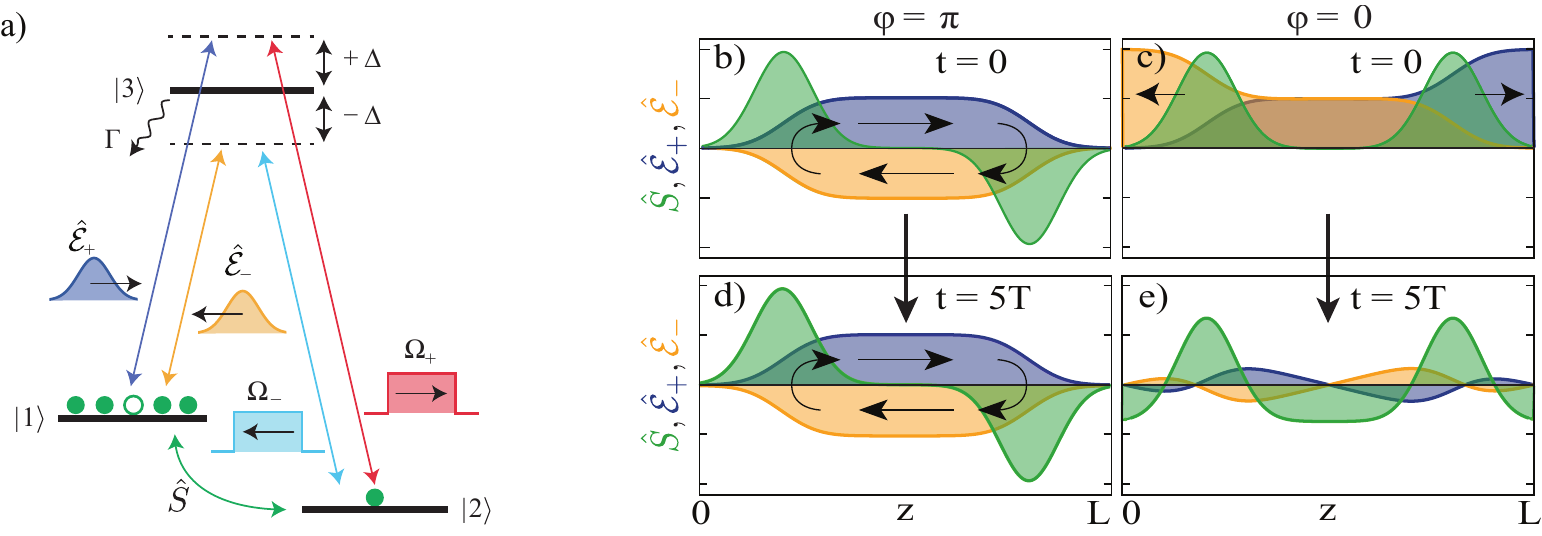}} 
\caption{\small (a) Level scheme for Raman SL. Forward (+) and backward (-) probe fields $\qE_\pm$ are coupled to the atomic spinwave $\hat{S}$ by corresponding control fields $\Omega_\pm$ with large equal and opposite single-photon detunings $\Delta_+ = \Delta$, $\Delta_- = -\Delta$. (b--e) Schematic of Raman SL formation. The initial spinwave consists of two identical separated Gaussian envelopes with equal or opposite phase. Simultaneous pumping by the two-colour control field produces SL. (b, d) The antisymmetric spinwave, $\phi = \pi \implies \int \mathrm{d}z \, \hat{\mathcal{S}} = 0$, is stationary and decays only globally---giving rise to a stable SL field. (c, e) The symmetric spinwave, $\phi = 0 \implies \int \mathrm{d}z \, \hat{\mathcal{S}} \ne 0$, evolves until it reaches a stationary configuration, leaking light fields forward and backward in the process. Figure modified from \cite{everett_dynamical_2016}   }
\label{fig:raman_combined}
\end{figure*}

To generate Raman SL a suitable spinwave is written into the ensemble via some initial probe pulse sequence. The spinwave is subsequently illuminated with counterpropagating control fields that have large equal and opposite detuning as shown in \Cref{fig:raman_combined}(a). The forward and backward control fields of Raman SL, like two-colour EIT-SL in \Cref{fig:schemes}(d), are mutually detuned. In this case, however, the detuning is so large that the interaction bandwidth is much narrower than the mutual detuning $\Delta_+ - \Delta_-$. This prevents the formation of HOCs no matter the atomic temperature.

Due to the pair of counterpropagating control fields, the spinwave is converted into forward- and backward-propagating probe fields. Stable SL will be generated when the counterpropagating components interfere destructively such that the optical field vanishes at the edges of the ensemble and no field escapes. The spinwaves generated by each probe field travelling in opposite directions within the ensemble also interfere destructively, cancelling out any evolution. The spinwave and probe fields are therefore stationary, and any bright optical standing wave in the probe fields is trapped. The condition for a stable Raman SL configuration is elegant in its simplicity: all that is required is that the integral of the spin wave over the ensemble is zero:
\begin{align}
\int \mathrm{d}z \, \hat{\mathcal{S}} = 0 \,.     
\end{align}
Any spinwave that integrates to zero along the length of the memory will evolve only by a global decay rate, we can refer to such a spinwave as `stationary'. This Raman SL condition is satisfied by any spinwave with an average amplitude of zero---including spinwaves that are spatially separate from the SL field they generate.

A schematic of a simple Raman SL configuration satisfying this condition is shown in \Cref{fig:raman_combined}(b). The spinwave consists of two equal, but separated, Gaussian coherence envelopes with opposite phase. The forward and backward probe fields driven from the spinwave by counterpropagating control fields interfere destructively at the ensemble edges, but between the two components of the spinwave exists a stationary optical field consisting of equal forward and backward fields with opposite phase circulating between the coherences. The spinwaves are stationary (up to a global decay rate) and evolve unchanged to \Cref{fig:raman_combined}(d). 

\begin{figure}[b!]
\centerline{\includegraphics[width=80mm]{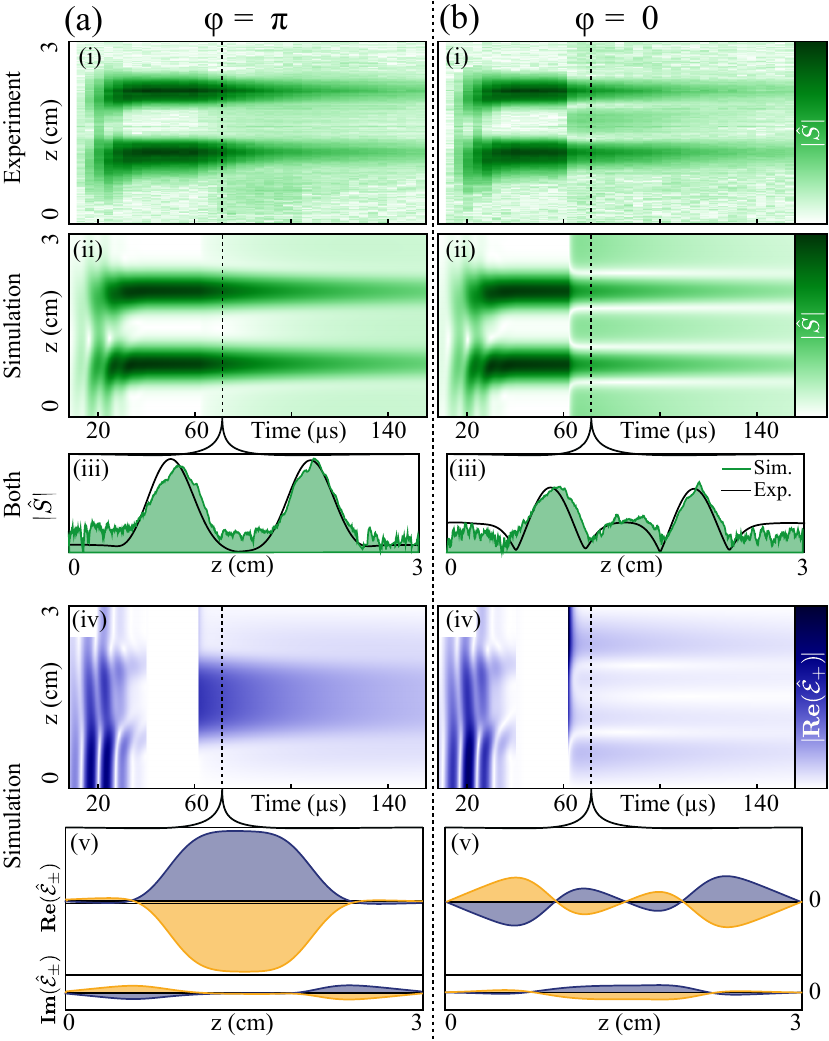}} 
\caption{\small (Adapted from \cite{everett_dynamical_2016}). A comparison of experimental results and modelling for Raman SL. (a) Stationary and (b) initially non-stationary spinwaves are illuminated by counterpropagating control fields at 60 $\mu$s. (i) Experimental imaging and (ii) modelling compare the evolution of the spinwaves, with (iii) a snapshot of both at 64 $\mu$s. (iv) the modelled forward propagating probe field. Intense light emerges from the non-stationary spinwave during its rapid evolution to a stationary spinwave. (v )A snapshot of both modelled probe fields at 64 $\mu$s, the trapped light causes no further evolution of the spinwave.}
\label{fig:raman_images}
\end{figure}

 If the initial spinwave has instead $\int \mathrm{d}z \, \hat{\mathcal{S}} \ne 0$ then $\hat{\mathcal{E}}^\pm\left(z=0,L \right) \ne 0$. In this case the probe fields evolve and leak from the ensemble until the spinwave reaches an equilibrium state with zero mean. At equilibrium, $\qE^\pm$ circulate within the ensemble and the resulting spinwave is distributed such that $\qE^\pm$ interfere destructively to arrest any further evolution. For example, by flipping the phase of one component of the spinwave in \Cref{fig:raman_combined}(b) such that the relative phase between the two Gaussians is $\phi=0$, we have a spinwave that is unstable under Raman SL. This spinwave, shown in \Cref{fig:raman_combined}(c), evolves over time into the stable spinwave shown in \Cref{fig:raman_combined}(e). A substantial proportion of the coherence may escape in this process.

Raman SL was first characterized in an ensemble of laser-cooled $^{87}$Rb atoms \cite{everett_dynamical_2016}. In this case the initial spinwave was written into the ensemble via a gradient-echo scheme \cite{alexander_photon_2006,hetet_photon_2008, longdell_analytic_2008}. By side-imaging the atomic coherence during the Raman SL process, the authors were able to compare the evolution of the symmetric and anti-symmetric spinwaves shown in \Cref{fig:raman_combined}(b,d) and (c,e). The corresponding results are shown in \Cref{fig:raman_images}(a) and (b). The magnitude of the spinwave is shown by the optical depth distribution along the optical propagation axis ($z$). This distribution changes in time as the spinwave first propagates into the ensemble then evolves under SL. Because the symmetric, \Cref{fig:raman_images}(a), and anti-symmetric ,\Cref{fig:raman_images}(b), spinwaves differ only by a phase, the two OD distributions are initially the same. At  $60$~$\mu$s, the Raman SL control fields are activated and the distributions begin to evolve. The symmetric spinwave develops a distinctive central coherence peak that is absent from the evolution of the antisymmetric spinwave. This difference is most evident in \Cref{fig:raman_images}(iii), which compares coherence distribution of the two cases after they have both reached a SL equilibrium. The shape of the antisymmetric spinwave is essentially stable under Raman SL, as expected.

The spinwave evolution was further confirmed by mapping the coherence back to a traveling probe field subsequent to the SL operation. The field recalled from the symmetric spinwave was both weaker and distorted consistent with the Raman SL model. No considerable field escaped the ensemble during the SL process with an antisymmetric spinwave. However, the evolution of the symmetric spinwave under counterpropagating control was accompanied by the detection of brief, intense probe fields escaping in both directions from the unstable Raman SL configuration. These fields lasted until the spinwave had reached the stable configuration in (b-iii). 

\Cref{fig:raman_images}(iv) and (v) show the corresponding SL fields $\hat{\mathcal{E}}_\pm$ from simulations. The symmetric case features a considerable SL field at the centre of the ensemble, between the two atomic coherence peaks. This is a distinctive feature of Raman SL. In contrast to the EIT polariton, the spinwave and stationary probe field need not coincide, and the SL field can be bright in regions where the spinwave is essentially zero. Raman-SL grants dynamic control over not only the optical bandgap, but also the SL distribution within the ensemble. This additional capability of Raman-SL increases the range of SL-mediated interactions that are possible. We return to this subject in \Cref{sec:apps}.

\section{Theory of stationary light}\label{sec:theo}
Having seen the range of SL experiments to date, we will delve now into a theoretical model of the atom-light interaction that gives rise to SL.  Along the way, different approximations allow the model to branch into treatments of the EIT and Raman conditions as well as dealing with the higher order coherences.

The conditions for which a weak probe field is converted into SL may occur when three- or four-level atomic systems are driven by bright, counter-propagating control fields. We begin our analysis by deriving the behaviour for a four-level double-$\Lambda$ system driven by two counter-propagating control fields, as shown in \Cref{fig:schemes}(e).

We assume that the probe fields are weak and propagate in opposite directions. They may be described by the operators \cite{PhysRevA.76.033805}:
\begin{align}
\hat{\textbf{E}}_{p+}(z)=&\boldsymbol{\epsilon_{p+}}\sqrt{\frac{\hbar\omega_{p+}}{4\pi c \epsilon_0A}}\int_{\omega_{p+}}\mkern-18mu \mathrm{d}\omega\left(\hat{a}_\omega e^{i\omega z/c}+\hat{a}^\dag_\omega e^{-i\omega z/c}\right)\nonumber\\
\hat{\textbf{E}}_{p-}(z)=&\boldsymbol{\epsilon_{p-}}\sqrt{\frac{\hbar\omega_{p-}}{4\pi c \epsilon_0A}}\int_{\omega_{p-}}\mkern-18mu \mathrm{d}\omega\left(\hat{a}_\omega e^{-i\omega z/c}+\hat{a}^\dag_\omega e^{i\omega z/c}\right)\nonumber
\end{align}

We treat the bright control fields as classical, with electric fields:
\begin{align}
\textbf{E}_{c+}(z)=&\boldsymbol{\epsilon_{c+}}\mathcal{E}_{c+}(t-z/c)\textrm{cos}[\omega_{c+}(t-z/c)],\nonumber\\
\textbf{E}_{c-}(z)=&\boldsymbol{\epsilon_{c-}}\mathcal{E}_{c-}(t+z/c)\textrm{cos}[\omega_{c-}(t+z/c)].
\end{align}
The subscripts refer to the field and and direction of travel: $p$ is for probe; $c$ is for control; $+$ is for forwards and $-$ is for backwards.  The unit polarization vectors are given by $\boldsymbol{\epsilon}$ while the slowly varying field envelopes are given by $\mathcal{E}(t-z/c)$. The cross-sectional area of the beam is $A$.

We assume that the weak fields each exist in a small bandwidth around a carrier frequency given by $\omega_{p+}=\omega_{13}+\Delta_+$, $\omega_{p-}=\omega_{14}+\Delta_-$, $\omega_{c+}=\omega_{23}+\Delta_+$, $\omega_{c-}=\omega_{24}+\Delta_-$ where the frequency difference of each probe-field $\omega_{p\pm}$ relative to the atomic transition frequencies $\omega_{13}$, $\omega_{14}$ include an independent detuning $\Delta_\pm$ from the excited states.

The light interacts with an ensemble of $N$ atoms over a length $L$. We can then write the interaction part of the Hamiltonian as
\begin{align}
&\hat{H}_\mathrm{INT}=-\hbar\sum^{N}_{n=1}\Bigg[\Omega_{c+}(t-z_n/c)e^{-i\omega_{c+}(t-z_n/c)}\hat{\sigma}^n_{32} \nonumber \\
&\phantom{\hat{H}_\mathrm{INT}=} +\Omega_{c-}(t+z_n/c)e^{-i\omega_{c-}(t+z_n/c)}\hat{\sigma}^n_{42}  \nonumber\\ 
&+g\left(\frac{L}{2\pi c}\right)^{1/2}\Bigg(\int_{\omega_{p+}} \mathrm{d}\omega\hat{a}_\omega e^{i\omega z/c}\hat{\sigma}^n_{31}  \\
& \phantom{+g\left(\frac{L}{2\pi c}\right)^{1/2}\Bigg(}+ \int_{\omega_{p-}}\mathrm{d}\omega\hat{a}_\omega e^{-i\omega z/c}\hat{\sigma}^n_{41}\Bigg)+ \textrm{H.c.}\Bigg], \nonumber
\end{align}
where H.c. is the Hermitian conjugate. The Rabi frequency associated with the forward control field is $\Omega_{c+} = \bra{3}(\mathbf{\hat{d}}_{23}\cdot\boldsymbol{\epsilon_{c+}})\ket{2}\mathcal{E}_{c+}/(2\hbar)$ where $\mathbf{\hat{d}}_{23}$ is the transition dipole operator for the $\ket{2}\rightarrow\ket{3}$ transition, with a similar expression for the backward control field. The coupling rate between the forward probe field and the $\ket{1}\rightarrow\ket{3}$ transition is 
\begin{align}
g = \bra{3}(\mathbf{\hat{d}}_{13}\cdot\boldsymbol{\epsilon_{p+}})\ket{1}\sqrt{\frac{\omega_{p+}}{2\hbar c \epsilon_0 A L}} \,,
\end{align}
and, for simplicity, we assume that the coupling rate for the backward propagating probe is identical.

\vspace{2.5cm}

The equations of motion are more useful if we define operators that vary slowly compared to the optical frequencies over space and time. This requires a large density of atoms to allow assumptions about weak fields to hold. We take a slice $\mathrm{d}z$ of the ensemble containing a large number of atoms $N_z \gg 1$ and define the following operators: 
\begingroup
\allowdisplaybreaks
\begin{align} 
\hat{\sigma}_{\mu\mu}(z,t)&= \frac{1}{N_z}\sum^{N_z}_n\hat{\sigma}^n_{\mu\mu}(t), \nonumber\\
\hat{\sigma}_{32}(z,t)&=\frac{1}{N_z}\sum^{N_z}_n\hat{\sigma}_{32}^n(t)e^{-i\omega_{c+}(t - z_n/c)}, \nonumber\\
\hat{\sigma}_{42}(z,t)&=\frac{1}{N_z}\sum^{N_z}_n\hat{\sigma}_{42}^n(t)e^{-i\omega_{c-}(t + z_n/c)}, \nonumber\\
\hat{\sigma}_{31}(z,t)&=\frac{1}{N_z}\sum^{N_z}_n\hat{\sigma}_{31}^n(t)e^{-i\omega_{p+}(t- z_n/c)}, \label{eq:slowoperators} \\
\hat{\sigma}_{41}(z,t)&=\frac{1}{N_z}\sum^{N_z}_n\hat{\sigma}_{41}^n(t)e^{-i\omega_{p-}(t+ z_n/c)}, \nonumber\\
\hat{\sigma}_{21\pm}(z,t)&=\frac{1}{N_z}\sum^{N_z}_n\hat{\sigma}_{21}^n(t)e^{-i(\omega_{p\pm}-\omega_{c\pm})(t\mp z_n/c)}, \nonumber\\
\hat{\mathcal{E}}_\pm(z,t)&=\sqrt{\frac{L}{2\pi c}}e^{i\omega_{p\pm}(t\mp z/c)}\int_{\omega_{p\pm}} \mathrm{d}\omega\hat{a}_\omega(t) e^{\pm i\omega z/c}. \nonumber
\end{align}
\endgroup
The multiplication by terms of the form $\exp(-i\omega(t-z/c))$ assigns a separate rotating frame to each operator. The $\pm$ subscripts for $\sigma_{21\pm}$ and $\hat{\mathcal{E}}_\pm$ indicate that the operator is slowly varying with respect to fields travelling in the $\pm z$ direction. 
The collective operators have commutators
\begin{align}
\left[\hat{\sigma}_{\mu\nu}(t),\hat{\sigma}_{\alpha\beta}(t)\right]&=\delta_{\nu\alpha}\hat{\sigma}_{\mu\beta}(t) - \delta_{\mu\beta}\hat{\sigma}_{\alpha\nu}(t),\nonumber\\
\left[\hat{\mathcal{E}}_\pm(t),\hat{\mathcal{E}}_\pm^\dag(t)\right]&=1. \nonumber
\end{align}

Substituting the slowly varying operators into $\hat{H}_\mathrm{INT}$ and including the energy terms for the separate light and atomic systems gives the complete Hamiltonian:
\vspace{2cm}
\begin{widetext}
\begin{align}
\hat{H}= &\int \mathrm{d}\omega\,\hbar\omega\hat{a}_\omega^\dag\hat{a}_\omega - \frac{\hbar\omega_{p+}}{L}\int_0^L \mathrm{d}z\,\hat{\mathcal{E}}_+^\dag\hat{\mathcal{E}}_+-\frac{\hbar\omega_{p-}}{L}\int_0^L \mathrm{d}z \, \hat{\mathcal{E}}_-^\dag\hat{\mathcal{E}}_- \label{eq:3levelcphamiltonian}\\&+ 
\int^L_0\mathrm{d}z\, \hbar \mathcal{N}(z) \times \Bigg[\Delta_+\hat{\sigma}_{33}+\Delta_-\hat{\sigma}_{44}-\Bigg(\Omega_{c+}(t-z/c)\hat\sigma_{32} + \Omega_{c-}(t+z/c)\hat{\sigma}_{42}
+g\Big(\hat{\mathcal{E}}_+\hat{\sigma}_{31}+\hat{\mathcal{E}}_-\hat{\sigma}_{41} \Big)+ \textrm{H.c.}\Bigg)\Bigg]\nonumber
\end{align}
\end{widetext}
\vspace{0cm}

\noindent where $\mathcal{N}(z)$ is the linear atomic density. The $(z,t)$ dependence of the operators is generally omitted for readability.

 To obtain compact equations of motion that yield insight into the dynamics, we make three assumptions. The first is that the probe fields are weak enough that almost all the atomic population resides in $\ket{1}$. Known as the pure-state approximation, this allows us to keep track of only the coherences, as $\hat{\sigma}_{11}\approx1$ and $\hat{\sigma}_{22}\approx\hat{\sigma}_{33}\approx\hat{\sigma}_{44}\approx0$. Additionally, we assume that the length of the ensemble L is short enough that the free-space propagation time for light to traverse the ensemble is much faster than any timescale of interest, $L/c \ll T$. Finally, we assume that the two $\Lambda$ transitions, one formed by the forward propagating fields and the other by the backward propagating fields, are phase-matched $k_{p+}-k_{c+}= k_{p-}-k_{c-}$ and have equal two-photon detunings  $\omega_{p+}-\omega_{c+}= \omega_{p-}-\omega_{c-}$. This allows us to write a single slowly-varying $\hat{\sigma}_{21}$ coherence operator rather than separating it into components that each couple to the forward or backward fields. With these assumptions, we find familiar Maxwell-Bloch equations \cite{PhysRevA.76.033805}, with an additional probe field and corresponding excited state coherence all interacting with a single spinwave:
\begin{align}
\partial_t\hat{\sigma}_{13}&= -(\Gamma+ i\Delta_+)\hat{\sigma}_{13}+i g \hat{\mathcal{E}}_+ + i\Omega_{c+}\hat{\sigma}_{12} \nonumber\\
\partial_t\hat{\sigma}_{14}&= -(\Gamma+ i\Delta_-)\hat{\sigma}_{14}+i g \hat{\mathcal{E}}_- + i\Omega_{c-}\hat{\sigma}_{12} \nonumber\\
\partial_t\hat{\sigma}_{12}&= -\gamma\hat{\sigma}_{12}  +i\Omega_{c+}^*\hat{\sigma}_{13}+i\Omega_{c-}^*\hat{\sigma}_{14} \nonumber\\
\partial_z\hat{\mathcal{E}}_+ &= \frac{ig\mathcal{N}(z)L}{c}\hat{\sigma}_{13} \nonumber\\
\partial_z\hat{\mathcal{E}}_- &= \frac{ig\mathcal{N}(z)L}{c}\hat{\sigma}_{14}. \label{eq:3levelcp}
\end{align}

The equations can be simplified further by writing them in terms of typical experimental parameters. The optical depth $d=g^2NL/(\Gamma c)$ is a standard parameter and is straightforward to measure experimentally. We substitute it into the equations of motion to remove the atom number $N$ and interaction strength $g$. To avoid having a spatially dependent optical depth, we scale the spatial coordinate to be normalised according to the atomic density $\xi(z) = \int_0^z \mathrm{d}z'\,\mathcal{N}(z')/N$.  To replace terms containing the atom number, the coherences are also renormalised: $\hat{S}=\sqrt{N}\hat{\sigma}_{12}$, $\hat{P}_+=\sqrt{N}\hat{\sigma}_{13}$, and $\hat{P}_-=\sqrt{N}\hat{\sigma}_{14}$. The probe field is also renormalised $\hat{\mathcal{E}}_\pm \rightarrow \sqrt{c/(L\Gamma)}\hat{\mathcal{E}}_\pm $, giving

\begin{align}
\partial_t\hat{P}_\pm&= -(\Gamma+ i\Delta_\pm)\hat{P}_\pm+i \sqrt{d}\Gamma\hat{\mathcal{E}}_\pm+ i\Omega_{c\pm}\hat{S} \label{eq:3levelcp:1}\\
\partial_t \hat{S}    &= -\gamma \hat{S} + i \Omega_{c+}^* \hat{P}_+ + i \Omega_{c-}^* \hat{P}_-\label{eq:3levelcp:2}\\
\pm\partial_\xi\hat{\mathcal{E}}_\pm &= i\sqrt{d}\hat{P}_\pm  \label{eq:3levelcp:3} 
\end{align}

\subsection{EIT stationary light}
The original proposal for SL used EIT with an additional counterpropagating control field. This form of SL arises from the equations of motion when we set both control fields on resonance: $\Delta_\pm=0$. We can make an adiabatic approximation by assuming that the excited state coherences vary slowly relative to the excited state decay rates $\partial_t \hat{P} \ll \Gamma$.

We then consider the equations in the spatial Fourier domain \cite{zimmer_dark-state_2008,campbell_direct_2017} by applying the transform $X(\xi,t)=\int{\mathrm{d}\kappa \, e^{-i\kappa\xi}\Tilde{X}(\kappa,t)}$ and expand to first order in $\kappa/d$. Combining \Cref{eq:3levelcp:1} and \Cref{eq:3levelcp:3}, we find
\begin{align}
    \Tilde{\hat{\mathcal{E}}}_\pm \simeq -\frac{\Omega_{c\pm}}{\sqrt{d}\Gamma}(1\pm i\kappa/d)\Tilde{\hat{S}} .
\end{align}
The quantity $\kappa/d$ describes the spatial variation of the coherence S with respect to the optical depth. After substituting this, along with \Cref{eq:3levelcp:1}, into \Cref{eq:3levelcp:2} we transform back into the the spatial domain $\xi$ to obtain the approximate equation of motion \cite{campbell_direct_2017}
\begin{align}
    \left(\partial_t + \Gamma \tan^2\theta\left(\cos 2\phi\partial_\xi-\frac{1}{d}\partial^2_\xi\right)\right)\hat{S}=0
\end{align}
with the mixing angles
$\tan^2\theta\equiv(|\Omega_+|^2+|\Omega_-|^2)/(d\Gamma^2)$ and $\tan^2\phi\equiv|\Omega_-|^2/|\Omega_+|^2$.

The diffusion term $\partial^2/\partial \xi^2$ is due to the finite absorption length of light in the EIT medium; the two probe fields become unequal where the coherence changes quickly in space and there is no longer complete interference between the two. The pulse matching is therefore imperfect, allowing decay of the polariton. For sufficiently large optical depth, the diffusion can be neglected in order to define a dark state polariton for the system:

\begin{align}
    \hat{\Psi}_D = \sin \theta(\hat{\mathcal{E}}_+\cos \phi + \hat{\mathcal{E}}_-\sin \phi) - \hat{S} \cos \theta
\end{align}

This dark state polariton can also be found as in Zimmer et al. \cite{zimmer_dark-state_2008} by applying a Morris-Shore transformation to the system. By analogy with the Schrodinger equation for a massive particle, the diffusion term for the dark-state polariton identifies a complex effective mass of the polariton:
\begin{align}
    m^* = 2\hbar\left(\frac{d\Gamma}{\Omega}\right)^2\frac{1}{\Delta-i\Gamma}
\end{align}
This mass is relevant in applications of SL for generating non-classical statistics of the polariton.

It is also possible to find a bandgap in the dispersion relation by analysing the equations of motion in the temporal frequency domain \cite{moiseev_stationary_2014}. Frequency domain analysis of SL is particularly important in proposals for quantum gates based on changes in dispersion for light transmitted or reflected from the ensemble \cite{lahad_induced_2017,iakoupov_controlled-phase_2018}.

\subsection{Raman stationary light}
As described in \Cref{sec:ORSL}, Raman SL relies on destructive interference between light emitted from different regions of the ensemble. To obtain a simple equation for this behaviour, we follow Ref.~\cite{everett_dynamical_2016} and take the two counterpropagating probe fields far from resonance. We make the secular approximation from the start and justify this later based on the large difference in detuning required in the probe fields. At large detuning $\Delta \gg \partial_t\hat{P}$, 

\begin{align}
    \hat{P}_\pm\approx i\left(\sqrt{d}\Gamma\hat{\mathcal{E}_\pm} + \Omega_{c\pm}\hat{S}\right)/\left(\Gamma+i\Delta_\pm\right)\label{eq:ramanadiabaticapprox}
\end{align}

We can ignore the incoherent absorption of the probe fields due to the excited state, but should still consider the dispersion. Substituting \Cref{eq:ramanadiabaticapprox} into \Cref{eq:3levelcp:3}, the probes experience phase rotations $\partial_\xi\hat{\mathcal{E}}_\pm\rightarrow i\Gamma d\hat{\mathcal{E}}_\pm/\Delta_\pm$. Loss from the spinwave due to incoherent absorption of the control field is collected along with spinwave dephasing in $\gamma$, resulting in:
\begin{align}
\partial_t\hat{S}=& -(\gamma-i\left(\frac{|\Omega_{c+}|^2}{\Delta_+}+\frac{|\Omega_{c-}|^2}{\Delta_-}\right)\hat{S}\nonumber  \\&+i\sqrt{d}\Gamma\left(\frac{\Omega_{c+}^*}{\Delta_+}\hat{\mathcal{E}}_++\frac{\Omega_{c-}^*}{\Delta_-}\hat{\mathcal{E}}_-\right)\label{eq:2levwdisp1}\\
\partial_\xi\hat{\mathcal{E}}_+ =& i\sqrt{d}\frac{\Omega_{c+}}{\Delta_+}\hat{S} + i\frac{\Gamma d}{\Delta_+}\hat{\mathcal{E}}_+ \label{eq:2levwdisp2}\\
\partial_\xi\hat{\mathcal{E}}_- =&- i\sqrt{d}\frac{\Omega_{c-}}{\Delta_-}\hat{S} - i\frac{\Gamma d}{\Delta_-}\hat{\mathcal{E}}_-.\label{eq:2levwdisp3}
\end{align}

\subsubsection*{The Raman stationary light equation \label{sec:sleqs}}
As with EIT SL, we set the control field drivings equal, $\Omega_{c+}=\Omega_{c-}=\Omega$. To achieve the interference effect that produces Raman SL, it is necessary to ensure that light generated in one location will interfere with a uniform phase throughout the ensemble. Where the dispersion terms above are non-negligible, this can be satisfied by setting equal and opposite  $\Delta_+=-\Delta_-=\Delta$. The dispersion term $i\Gamma d/\Delta$ is now equal for both probe fields, and can be removed by transforming to the rotating spatial frame:
\begin{align}
\hat{\mathcal{E}}_\pm\rightarrow\hat{\mathcal{E}}_\pm e^{(i\frac{\Gamma d}{\Delta} \xi)},
\end{align}
giving
\begin{align}\label{eq:cptwoleveleqns3}
\partial_t\hat{S}(t,\xi)&=i\sqrt{d}\,\frac{\Gamma \Omega}{\Delta}\left(\hat{\mathcal{E}}_++\hat{\mathcal{E}}_-\right)-\gamma\hat{S}\\\label{eq:cptwoleveleqns4}
\partial_\xi\hat{\mathcal{E}}_+(t,\xi)&=  i\sqrt{d}\frac{\Omega}{\Delta}\hat{S}\\
\partial_\xi\hat{\mathcal{E}}_-(t,\xi)&= - i\sqrt{d}\frac{\Omega}{\Delta}\hat{S}\label{eq:cptwoleveleqns5}.
\end{align}

Solutions of the probe field are found by integrating the spinwave:
\begin{align}
\hat{\mathcal{E}}_+(t,\xi)&=i\sqrt{d}\frac{\Omega}{\Delta}\int_0^\xi\hat{S}(t,\xi') \, \mathrm{d} \xi' \label{eq:forwardprobeintegral}\\
\hat{\mathcal{E}}_-(t,\xi)&=-i\sqrt{d}\frac{\Omega}{\Delta}\int_1^\xi\hat{S}(t,\xi')\,\mathrm{d}\xi'.\label{eq:backwardprobeintegral}
\end{align}
The SL equation is then obtained by inserting these solutions into \Cref{eq:cptwoleveleqns3}:
\begin{align}
 \left( \partial_t + \gamma\right) \hat{S}(t,\xi) 
= -d\,\Gamma\frac{\Omega^2}{\Delta^2}\int_0^1{\hat{S}(t,\xi') \, \mathrm{d}\xi'} \,.\label{eq:sleq}
\end{align}
Apart from uniform decay, a spinwave that integrates to zero along the length of the ensemble will not evolve. 

This equation can also be used to describe the evolution of spinwaves that do not satisfy $\int{S} \, \mathrm{d}\xi =0$. The spinwave can be separated into a spatially constant term and a part that integrates to zero. In other words, the spinwave can be written as the sum of a stationary and non-stationary component $\hat{S}(\xi,t)=\hat{S}_\xi(\xi) + \hat{S}_t(t)$ where $\int_0^1\hat{S}_\xi(\xi') \, \mathrm{d}\xi'=0$. The SL equation can be applied at time $t=0$.
\begin{align}
(\partial{t}+\gamma)\hat{S}(t,\xi)&=-d\frac{\Gamma\Omega^2}{\Delta^2}\int_0^1\hat{S}(t,\xi')\, \mathrm{d}\xi'\\
& =-d\frac{\Gamma\Omega^2}{\Delta^2}\hat{S}_t(t)
\end{align}

By linearity, the stationary component will remain constant and the non-stationary component will decay exponentially:
\begin{align}
\hat{S}(\xi,t)=[\hat{S}_{\xi}(\xi) + \hat{S}_{t_0}e^{-t(\frac{d\Gamma\Omega^2}{\Delta^2})}]e^{-\gamma t}
\end{align}

\subsection{Higher order coherences}
We have so far considered the secular case; where the two probe fields address entirely separate excited states, or implicitly made a secular approximation. In \Cref{sec:hocs} we discussed higher order coherences (HOCs), and now we provide some mathematical tools for modelling this phenomenon.

In the non-secular case, the same excited state is addressed by the counterpropagating fields, allowing additional couplings between the fields. A high momentum state in the atomic coherence, or HOC, is generated when a probe or control photon is absorbed and re-emitted into the control field travelling in the opposite direction. The short wavelength causes HOCs to decay more quickly in non-stationary media, and the momentum mismatch means they do not couple equally to both probe fields, effecting the SL behaviour.

Due to the shorter wavelengths the HOCs are not described by the same slowly varying operators. Instead additional operators can be used to describe a spinwave with HOC components. For example Wu et al. \cite{wu_stationary_2010} write the spinwave as a sum of operators,
\begin{align}
\hat{\sigma}_{12}=\sum_{n=-\infty}^\infty \hat{\sigma}_{12}^{(2n)}e^{in[(\omega_{c+}+\omega_{c-})z/c+(\omega_{c+}-\omega_{c-})t]}
\end{align}
where each additional term is generated by absorbing a control photon travelling in one direction and emitting it in the other. 

The excited state coherences then coupling between each of the spinwave terms are
\begin{widetext}
\begin{align}
\hat{\sigma}_{13}&=e^{-i\Delta_+t+i\omega_{c+}z/c}\sum_{n=0}^\infty \hat{\sigma}_{13}^{(2n+1)}e^{in[(\omega_{c+}+\omega_{c-})z/c+(\omega_{c+}-\omega_{c-})t]}\\&+e^{-i\Delta_-t-i\omega_{c-}z/c}\sum_{n=0}^{-\infty} \hat{\sigma}_{13}^{(2n-1)}e^{in[(\omega_{c+}+\omega_{c-})z/c+(\omega_{c+}-\omega_{c-})t]}.
\end{align}

In the case of a standing wave control field $\omega_{c+}=\omega_{c-}$ the coupling between the spinwaves depends simply on the coupling of each control field with the relevant spinwave, giving equations of motion
\begin{align}
\partial_t\hat{\sigma}_{13}^{(\pm1)} &= -(\Gamma-i\Delta)\hat{\sigma}_{13}^{(\pm1)}+i\sqrt{d}\Gamma\hat{\mathcal{E}}_\pm+i\Omega_{c\pm}\hat{\sigma}_{12}^{(0)}+i\Omega_{c\mp}\hat{\sigma}_{12}^{(\pm2)}\\
\partial_t\hat{\sigma}_{13}^{(\pm(2n-1))} &= -(\Gamma_n-i\Delta)\hat{\sigma}_{13}^{(\pm1)}+i\Omega_{c\pm}\hat{\sigma}_{12}^{(\pm(2n-2))}+i\Omega_{c\mp}\hat{\sigma}_{12}^{(\pm2n)}\\
\partial_t\hat{\sigma}_{12}^{(0)} &=-\gamma\hat{\sigma}_{12}^{(0)} + i\Omega^*_{c+}\hat{\sigma}_{13}^{(+1)} + i\Omega^*_{c-}\hat{\sigma}_{13}^{(-1)}\\
\partial_t\hat{\sigma}_{12}^{(\pm2n)} &= -\gamma_{n}\hat{\sigma}_{12}^{(2n)} + i\Omega^*_{c+}\hat{\sigma}_{13}^{(2n+1)} + i\Omega^*_{c-}\hat{\sigma}_{13}^{(2n-1)}
\end{align}
\end{widetext}

This involves coupling along an infinite ladder of coherences, but the equations can be solved by truncating at a suitable order. Except in completely stationary atoms, the motional decay $\gamma_n$ of the higher order spinwaves is fast enough that only a few terms should be considered.

We have assumed no spatial ordering of the atoms thus far. In the case of standing wave control fields interacting with spatially ordered ultracold or stationary emitters, the dispersion relation becomes more complicated. The analytic results in Ref.~\cite{iakoupov_dispersion_2016} are beyond the scope of this review, but that work is also interesting for studying SL with disordered atoms for its independent derivation of the dispersion relations for both these cases.

\subsection{Phase matching and transverse propagation}\label{sec:phasematch}
Phase matching is an important concept for any light-matter interaction where additional optical frequencies are generated or directions of propagation change. Ideal phase matching means that all the light generated in a spatially extended interaction interferes constructively with the propagating field. Phase matching is critical for generating both EIT SL and Raman SL fields, but was taken for granted in our earlier derivation of the SL equations. 

Mathematically, phase matching for stationary light means orchestrating the properties of the optical fields so that the spinwave operators in \Cref{eq:slowoperators} are equal, i.e. $\sigma_{21+}=\sigma_{21-}$. This condition is satisfied naturally with probe and control pairs of equal frequency. Since this is not possible in many of the SL schemes we have discussed, it is necessary to match the phases by introducing an additional degree of freedom, namely by allowing for control fields that are not exactly parallel to their corresponding probe fields. We can write the spinwave operators with the longitudinal ($z$) component of the field momenta to give a phase factor

\begin{align}
    e^{-i[(\omega_{p\pm}-\omega_{c\pm})t+(k_{pz\pm}-k_{cz\pm})z]}.
\end{align}

Choosing a level scheme where the control fields have a larger momentum than the probe fields will allow $(k_{pz+}-k_{cz+})z=(k_{pz-}-k_{cz-})z=0$ by introducing a small angle between probe and control fields. This is illustrated, for example, in \Cref{fig:EITSL_imaging}(a), which shows the phase matching scheme used in Ref.~\cite{campbell_direct_2017}.

Under EIT conditions, applying control fields that do not satisfy phase matching can cause significant loss, as the interference causing the transparency breaks down. This would be detrimental to EIT SL. This loss mechanism does not exist in Raman SL due to the fields being far detuned from resonance. In this case, the interference condition that allows Raman SL  can only hold for all points in space with proper phase matching of the two counterpropagating probe fields generated from the spinwave. Without this, the Raman SL condition will not be satisfied leading to spinwave decay as the probe field leaks from the atomic ensemble. 

The theory collected here considers only plane wave propagation. This is acceptable for situations where the beam sizes are large enough, but in any other system the transverse mode could be relevent. Andr{\'e} et al. \cite{andre_nonlinear_2005} point out that for EIT SL in  free-space ensembles, the intensity profiles of the control fields effectively produce a waveguide due to the spatially-varying refractive index. It was also seen in the results of the hollow-core experiments \cite{blatt_stationary_2016} that the transverse mode of the control fields gave rise to extra inhomogeneous broadening. The impact of the spatial mode on Raman SL has not yet been investigated in any detail.

\section{Applications of stationary light}\label{sec:apps}
The physics of SL is rich with complexity. The ability to engineer dispersion in these systems allows the simulation of numerous quantum phenomona such as Cooper pairing with photons, spin-charge separation and relativistic quantum field theories. The review by Noh and Angelakis~\cite{noh_quantum_2017} summarizes recent developments in this area. The primary technological application for SL is in the development of nonlinear phase gates in optical quantum information processing. This will be the focus of our attention in this section.

\subsection{Optical gates}
Deterministic quantum computing schemes require a nonlinear element in order to realise a universal set of operations. Ideally, one would like to build a \emph{cnot} gate where the presence of a single control photon will invert the phase of a target photon \cite{chuang_simple_1995,OBrien1567}. This corresponds to a cross phase modulation so strong as to give a $\pi$ phase shift at the single photon level.  This feat has recently been achieved using a single atom in an optical resonator \cite{Hacker:wr} and an ensemble of Rydberg atoms \cite{Tiarks:2018tl}.  Theoretical work has suggested a way around needing a $\pi$ phase shift \cite{Munro:2005hk}.  In this scheme, phase shifts on the order of milliradians could be used to build deterministic optical gates using a combination of single photon and coherent states, although even milliradian single-photon phase shifts remain challenging.

 SL is a means of enhancing the available phase shifts simply by increasing the available interaction time. This could enable the construction of nonlinear phase gates in systems that would otherwise be too weakly interacting for use in any computation. SL has further promise in modifying the propagation of light during the nonlinear interaction to avoid parasitic effects limiting the fidelity of the operation.

\begin{figure*}[t!]
\includegraphics[width=175mm]{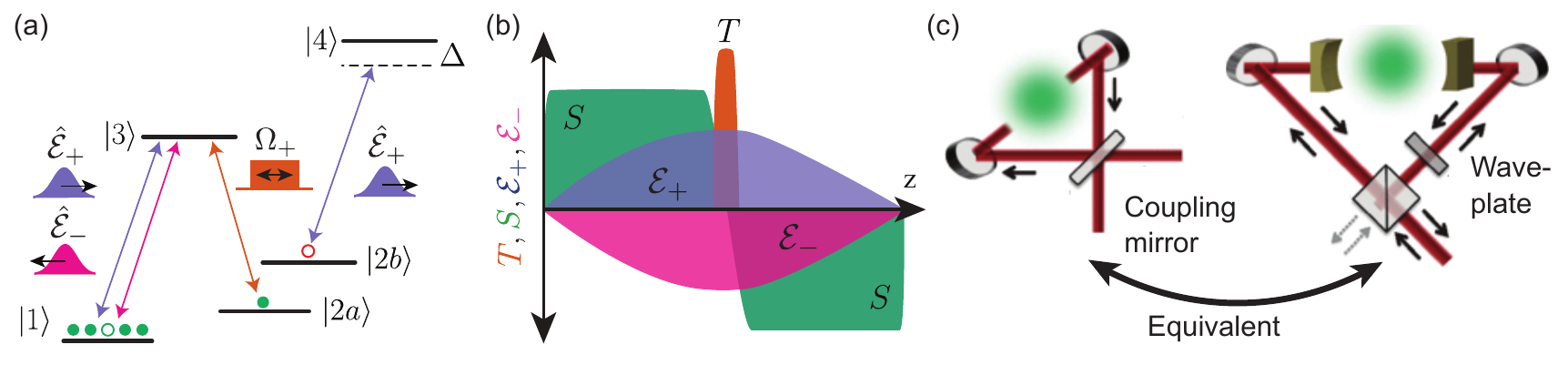}
\caption{  \small (a) Level scheme for cross nonlinearities with SL. One or both SL fields interact with a second level scheme, experiencing dispersion and/or modulating the energy level of the additional transition. (b) Arrangement of spinwave and fields allowing a control phase gate to be implmented with Raman SL  (c) Lahad and Firstenberg \cite{lahad_induced_2017} compare the transmission through the SL medium within a ring cavity to transmission through a Fabry-Perot cavity.}
\label{fig:combinedgates}
\end{figure*}

Andr\'e et al. \cite{andre_nonlinear_2005} proposed storing a weak pulse in an atomic ensemble and passing a second pulse across it in a quasi-SL configuration with unbalanced counterpropagating control fields. Friedler et al. \cite{friedler_deterministic_2005} proposed a similar scheme in which a slow light pulse travels through a SL pulse that modifies the two-photon detuning of the slow light pulse by AC-Stark shift and produces a phase shift. A typical level scheme for these schemes is shown in \Cref{fig:combinedgates}(a). Both these schemes were calculated to produce conditional $\pi$ phase shifts between single photons in experimental conditions that are currently accessible.

An in-principle demonstration of such a scheme was performed by Chen et al. \cite{chen_demonstration_2012}. A SL pulse was used to erase a weak stored pulse, rather than to imprint a phase shift, by a resonant interaction with atoms in the stored pulse spinwave. This showed usefully large phase shifts of up to 10~mrad could be achieved with this type of interaction.

After these first proposals, enhanced Kerr nonlinearities in EIT, including EIT SL, were the subject of several papers arguing that producing large phase shifts with such schemes was either impossible or extremely impractical \cite{shapiro_single-photon_2006,gea-banacloche_impossibility_2010,he_transverse_2011,he_continuous-mode_2012}. The diverse nature of the proposals makes the no-go theorems difficult to generalise, but a recent work by Viswanathan and Gea-Banacloche \cite{viswanathan_analytical_2018} summarises the obstacles as follows. There are two mechanisms which reduce the fidelity: phase noise arising from a finite response time of the medium; and entanglement arising from the phase-shifting interaction, in which photons are destroyed and recreated subject only to energy conservation. A theme of the no-go papers is that these mechanisms cannot be eliminated simultaneously while generating a large phase shift.

Recently, proposals for avoiding these pitfalls have begun to emerge, with proposals based on SL taking advantage of the radically different propagation of light compared with EIT schemes. Iakoupov et al. \cite{iakoupov_controlled-phase_2018} and Lahad and Firstenberg \cite{lahad_induced_2017} proposed sending light at frequencies outside the bandgap, within a transmission peak as shown in \Cref{fig:bandgap}. These transmission resonances exist wherever the polariton forms a standing wave inside the ensemble \cite{iakoupov_controlled-phase_2018}, and a constructive interference results for light at the far end of the ensemble. The propagation of light resembles the reflection of light within an optical cavity, lending the term \textit{induced cavity}. The multiple reflections of the light travelling through the ensemble change the character of the phase-shifting interaction and allow a high-fidelity, large phase shift. Lahad and Firstenberg make explicit comparison to a cavity as shown in \Cref{fig:combinedgates}(c)

Murray and Pohl proposed a slightly different approach based on a Rydberg interaction. A probe photon is incident on an ensemble illuminated by counterpropagating control fields. An additional coupling between a forward propagating probe and a Rydberg level prevents the forward propagating probe light from coupling to backward travelling probe light and vice versa. The probe photon is transmitted under EIT conditions. A stored `gate' photon shifts the Rydberg level for nearby atoms, interrupting the additional coupling and restoring SL conditions. The probe field then experiences a bandgap and is reflected \cite{murray_coherent_2017}.

Everett et al.~\cite{everett_dynamical_2016} proposed that a gate based on Raman-SL, shown in \Cref{fig:combinedgates} (a) and (b) would also overcome obstacles to high fidelity gates. SL is generated by a spinwave that is spatially separated from the nonlinear interaction of that light with a target state. The circulation of light from the spinwave through the interaction region is equivalent to routing light through an interaction region many times, for example by using a cavity. The combination of spatial separation and repeated weak interaction is proposed to escape the entanglement and phase noise problems.

These recent proposals all include the use of reflection to change the character of the nonlinear interaction. More work needs to be done to understand how and to what extent the no-go theorems are addressed by these schemes.

\section{Conclusion and outlook}
The ability to make states of SL with a group velocity of zero is not only inherently fascinating, it is also a useful technique with numerous applications. Experiments have now been carried out using hot and cold atomic vapours, as well as in hollow-cored fibres. The early understanding of SL arising from standing waves of the control field was shown to be incomplete and demonstrations of SL without standing waves have solidified our understanding of the phenomenon via a multi-wave mixing model.
In this review, our goal was to put all the demonstrations to date in context with a unifying theoretical model that shows how SL based on EIT and Raman interactions can be understood, as well as the behaviour of these schemes in hot and cold atomic systems where higher-order coherences may play a role.  In the future we look forward to further application and demonstrations of SL for quantum simulations and development of quantum information systems.

\subsection{Acknowledgments}
This work was supported by the Australian Research Council Centre of Excellence Grant No. CE170100012.

\bibliographystyle{ieeetr} 
\bibliography{SLrevesc}

\end{document}